# Design of a sustainable pre-polarizing magnetic resonance imaging system for infant hydrocephalus

**Johnes Obungoloch**[1,2,3], **Joshua Harper**[1,4], **Steven Consevage**[1,5], **Igor Savukov**[6], **Thomas Neuberger**[2,7], **Srinivas Tadigadapa**[8], **and Steven J. Schiff**[1,2,4, 5, 9]

[1]Center for Neural Engineering, The Pennsylvania State University, University Park, 16802 USA

[2]Department of Biomedical Engineering, The Pennsylvania State University, University Park, 16802 USA

[3]Mbarara University of Science and Technology, P.O Box 1410, Mbarara, Uganda

[4]Department of Engineering Science and Mechanics, The Pennsylvania State University, University Park, 16802 USA

[5]Department of Physics, The Pennsylvania State University, University Park, 16802 USA

[6]Los Alamos National Laboratory, New Mexico

[7]The Huck Institutes of the life Sciences, The Pennsylvania State University, University Park, 16802 USA

[8]Department of Electrical Engineering, The Pennsylvania State University, University Park, 16802 USA

[9]Department of Neurosurgery, The Pennsylvania State University, University Park, 16802 USA

**Corresponding author**
Dr. Steven J. Schiff
Email: sschiff@psu.edu
Phone: +1 814 863-4210, +1 814 865-2481

**ABSTRACT**

The need for affordable and appropriate medical technologies for developing countries continue to rise as challenges such as inadequate energy supply, limited technical expertise and poor infrastructure persists. Low-field MRI is a technology that can be tailored to meet specific imaging needs within such countries. Its low power requirements and the possibility of operating in minimally shielded or unshielded environments make it especially attractive. Although the technology has been widely demonstrated over several decades, it is yet to be shown that it can be diagnostic and improve patient outcomes in clinical applications. We here demonstrate the robustness of pre-polarizing MRI (PMRI) technology for assembly and deployment in developing countries for the specific application to infant hydrocephalus. Hydrocephalus treatment planning and management requires modest spatial resolution, and only that the brain can be distinguished from fluid – tissue contrast detail within the brain parenchyma is not essential. We constructed an internally shielded PMRI system based on the Lee-Whiting coil system with a 22 cm diameter of spherical volume. In an unshielded room, projection phantom images were acquired at 113 kHz with in-plane resolution of 3 mm x 3 mm, by introducing gradient fields of sufficient magnitude to dominate the 5000ppm inhomogeneity of the readout field. The low cost, straightforward assembly, deployment potential, and maintenance requirements demonstrate the suitability of our PMRI system for developing countries. Further improvement in the image spatial resolution and contrast of low-field MRI will broaden its potential clinical utility beyond hydrocephalus.

**Introduction**

Hydrocephalus is the most common condition in children world-wide that requires brain imaging and neurosurgical treatment. It is estimated that there are as many as 200,000 new cases of hydrocephalus per year throughout sub-Saharan Africa (SSA) alone [1]. In SSA, the majority of cases can be postinfectious, which increases the need for structural imaging to guide treatment since the bacterial brain infections can lead to substantially increased complexity of the fluid spaces within the brain [2]. The cost-effectiveness of treating infant hydrocephalus in SSA, with appropriate imagery, has been well shown [3]. Furthermore, it has recently been demonstrated in African infants that measuring brain volume on imaging can provide useful metrics of brain growth arrest with increased intracranial pressure prior to treatment, and catch-up growth in infants who do well cognitively following successful relief of pressure after surgery [4].

Hydrocephalus in infancy also presents one of the most straightforward challenges for magnetic resonance imaging (MRI). In these cases, the most important task is to segment an image into brain and cerebrospinal fluid (CSF). Rooney et al found that at 0.2 T, the T1 for CSF, white matter and grey matter are respectively 4.42s, 361ms, and 636ms [5]. This makes segmentation between brain tissue and CSF straightforward. The requirements for tissue contrast within the brain, or grey and white matter discrimination, are not important for diagnostics and treatment. Similarly, the spatial resolution required to affect treatment decisions are low – fluid compartments of at least multiple cubic centimeters in volume must be identified for fenestration or drainage. Although almost all new cases are seen in infancy, the heads can readily grow to adult size early within the first year of life. Fortuitously, the small infant body permits designs of imaging systems far smaller than that required for adult imaging.

MRI is arguably the safest technology for brain imaging that could be used in the diagnosis of hydrocephalus, but it is also the most expensive structural brain imaging modality available. Conventional high field (HF) MRI uses very strong magnetic fields generated by super-conducting magnets. Additionally, fringe fields from HF MRI systems must be shielded and this provides additional cost for the system. The high cost, which are typically $1 million USD per Tesla for whole body cryogenic MRI scanners [6], high power requirements, stringent technical demands and strict infrastructure specifications have hindered the proliferation of HF MRI in developing countries. As a less expensive option, developing countries more often use computed tomography (CT) for brain imaging [7]. While this alleviates many of the financial and infrastructure difficulties of HF MRI, CT contains components that are expensive to service and maintain in the developing world, and delivers relatively high doses of ionizing radiation, of particular hazard to young infants [8]. It is our position that the high tissue contrast and fine spatial resolution provided by typical CT or HF MRI substantially exceed what is required to reach surgical management decisions and achieve good patient outcomes in hydrocephalus.

Low-field (LF) coil-based MRI systems can offer an affordable, sustainable, and safe imaging alternative to HF MRI and CT for brain imaging in developing countries by reducing costs in both materials and manufacturing, lowering power requirements, eliminating the need for specialized equipment siting, and simplifying the technical aspects of troubleshooting, operation, and repair. The practicality of LF MRI was recognized from the inception of MRI technology in the 1980s. In 1985 for example, Sepponen et al. studied cerebral lesions using static LF MRI with a main field (Bo) of 20 mT and was able to establish considerable contrast between lesions and other brain tissues [9]. Nascimento et al. demonstrated whole-brain imagery with a low-cost system at 16 mT using a novel digital transmit-receive system [10]. The introduction of pre-polarizing MRI (PMRI) technology by Conolly and colleagues in 1993 expanded the potential benefits offered by LF MRI technology [11]. The PMRI uses two separately optimized electromagnet systems to provide a weak static and homogeneous magnetic field ($B_m$) at which the image is acquired, and a relatively strong pulsating magnetic field ($B_p$) to provide additional polarization before image acquisition. This technique allows acquisition of MR signal at very low frequencies provided by the $B_m$ field while taking advantage of the high signal to noise ratio (SNR) provided by the $B_p$ field. Other advantages of this technology include low power requirements, and relaxation of the magnetic field homogeneity requirements for the stronger $B_p$ field [11]. Using this technique, Conolly et al. acquired an image of a human hand with remarkable resolution and contrast using a $B_m$ field of 57 mT and Bp field of 0.4 T [12]. Similarly, with a reduced field strength of 2 mT for $B_m$ and a pre-polarizing field of 0.1 T, Savukov et al. performed imaging of the human hand [13], and partial imaging of the adult head [14].

PMRI is not the only low-field MRI technology that could be appropriate for developing countries, and it has its challenges. Rapidly ramping up the $B_p$ is practically challenging as it requires high intermittent power and introduces eddy currents into the system. Furthermore, there is strong induction coupling between the $B_p$ and $B_m$ which can negate the advantage gained by using $B_p$. Therefore, other methods of LF and sustainable imaging are under development as well. Recently, there have been demonstrations of imaging using static low-field MRI systems that also show promise. Lother et al. developed a novel design of a Helmholtz coil-based MRI system in which steel plates were used to boost the magnetic field within the coil arrangement and simultaneously shield the system from electromagnetic interference [15]. With field strength of 23 mT and a field of view (FOV) of 10 cm diameter of spherical volume (DSV), an image with in-plane resolution of 1.6 mm x 1.6 mm was acquired in 19 minutes. The system operates at 500 Watts. Similarly, Sarracanie et al. used a novel Helmholtz coil based LF system to acquire images of the human brain [16]. Working with a 6.5 mT main magnetic field generated using 220 cm diameter Helmholtz coils, images were acquired in 6 minutes with in-plane resolution of 2.5 mm x 3.5 mm.

Permanent magnet MRI systems, particularly Halbach arrays, are another type of LF MRI with possible applications in developing countries. Halbach arrays were introduced in 1980 when Klaus Halbach developed a method of cylindrically arranging small sized permanent magnets to generate uniform magnetic fields [17]. This method allows for construction of relatively low-weight permanent magnet MRI systems. Kimura et al. developed a Halbach array MRI system and used it for outdoor imaging of tree branches in 2011 [18]. However, because of the difficulty of creating gradient magnetic fields, mechanical rotation and translation of the system was required for slice selection and imaging. Meanwhile, Kose and Haishi were able to add gradient magnetic fields made of electromagnets to a permanent magnet system and acquired 2D images without the need for rotation and translation [19]. More recently, Cooley et al. used a Halbach array to acquire 2D phantom images without the use of gradients [20]. Their Halbach system had field strength of 77.3 mT and 16 cm field of view (FOV).

While Halbach array systems are attractive because of their low power requirement, it is challenging to generate uniform magnetic fields in large-bore sized systems [21]. Additionally, slice selection and 3D imaging using Halbach arrays compounds the design complexity by requiring mechanical rotations and translations or powered electromagnets to produce gradient fields. Static magnet LF MRI systems on the other hand require more continuous power than PMRI systems. Helmholtz design of static magnet MRI systems are especially power demanding. To generate a homogeneous magnetic field over the
FOV large enough to accommodate an adult head, the Helmholtz coils have to be at least several meters in diameter [22]. The current required to generate a magnetic field of 1 mT at the center of such a coil system would be 1100 Ampere-turns. The same field could be generated using Lee-Whiting coil arrangement with diameter of 0.5 m requiring only 28 Ampere-turns [23]. Furthermore, for very low readout fields, the SNR of PMRI is proportional to the product of $B_p$ and the square root of $B_m$ fields. For relatively higher readout fields above 23mT, the SNR becomes proportional to only $B_p$. Therefore, the SNR of a PMRI system can be comparable to that of static MRI systems where $B_o$ is equal to $B_p$ [11], provided the inductive effects of $B_p$ coil switching are eliminated. In this way, generating a small static magnetic field ($B_m$) while periodically switching the stronger magnetic field ($B_p$) saves power and mitigates heating, while matching the SNR capabilities of static field LF MRI systems. For these reasons, we have chosen to design a PMRI system for our introduction into a developing country context. We modified and re-designed the PMRI system from [14] to accommodate a 22 cm DSV FOV and operate at 550 W. The $B_m$ field was generated by a Lee-Whiting coil arrangement, the $B_p$ field was generated by a solenoid while the longitudinal gradient field was generated by a Maxwell coil pair and the transverse gradients by Golay coils.

The RF coils were saddles capable of accommodating an adult human head. We demonstrated the feasibility of using this PMRI system by imaging a variety of test phantoms with only internal shielding within the device.

**Materials and methods**

**PMRI coils**
The coil configuration of the PMRI system includes the $B_m$ coil, $B_p$ coil, gradient coils and radio frequency coils. These are described in the methodology below.

**The $B_m$ coil**
In designing the $B_m$ coil, homogeneity of the magnetic field is a major factor to consider. Magnetic field inhomogeneities introduce artifacts that degrade the quality of the resulting image. A number of coil designs including the Merrit three, four and five coil systems [24], the Rubens five coil system [25], Helmholtz and Lee-Whiting coil system [23] among others have been shown to generate acceptable field homogeneities. Of these, the Lee-Whiting coil system is relatively easy to implement and produces sufficiently homogeneous magnetic fields over large volumes [26]. Magnetic field homogeneity requirements for low field MRI systems have varied between researchers, but Lother et al was able to generate acceptable images with magnetic field homogeneities of 1200 ppm over a 10 cm DSV [15].

The Lee-Whiting coil system consists of two coil pairs symmetrically positioned from the center along the z-axis [23], with Ampere turns ratios of 9:4:4:9. Using the Biot-Savart law, the magnetic field $B_z$ along the axis of the Lee-Whiting coil system is given by

$$B_z = \frac{9\mu_o n I a^2}{2(a^2+(z-s_1))^{3/2}} + \frac{4\mu_o n I a^2}{2(a^2+(z-s_2))^{3/2}} + \frac{4\mu_o n I a^2}{2(a^2+(z+s_2))^{3/2}} + \frac{9\mu_o n I a^2}{2(a^2+(z+s_1))^{3/2}} \qquad \text{Equation 1}$$

where $n$ is the number of turns, $a$ is the radius of the coil system, $I$ is the current in the coils, $s_1$ and $s_2$ are positions of the inner and outer coils respectively from the center along the axis of the coil arrangement, and $\mu_0$ is permeability of free space. The field at the center ($z = 0$) is $\approx 17.96/D$ μT/A for a single turn [23], where $D$ is the diameter of the coil arrangement.

In our design of the $B_m$ system, a power requirement of 500 W at 10 A and 5 Ohm resistance was imposed as a design constraint. Coil selection was based on power restrictions and wire size. The wire length was determined by the resistance.

Enameled copper wire with nominal diameter of 2.07 mm and resistance of 2.315 x $10^{-3}$ Ω/m was used [27]. The length of the conductor was calculated to be 940 meters. The Lee-Whiting coil system was then wound with outer coils having 207 turns and the inner coils having 92 turns. The diameter of the Lee-Whiting coil system was 47.5cm. The axial separation of the inner and outer coils were ± 5.8cm and ± 22.3cm respectively from the center of the coil system.

COMSOL, a finite element analysis and solver software, was used to simulate the theoretical level of magnetic field homogeneity achievable for the given dimensions of the coil. COMSOL calculates the map of the static magnetic field using the Ampere's equation

$$\nabla \times H = J \qquad \text{Equation 2}$$

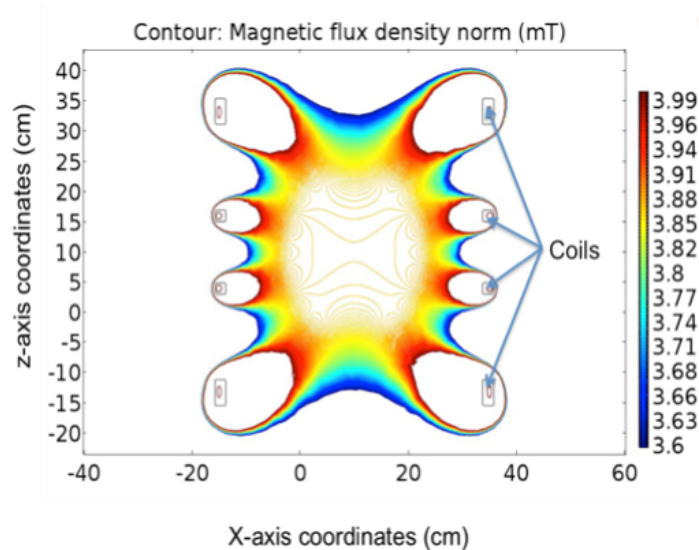

Figure 1: COMSOL simulation contour plot of the $B_m$ field homogeneity on the center xz-plane. The inner coils of the Lee-Whiting system had 92 turns each and the outer coils had 207 turns each. A direct current of 4.71 A was injected into the coils. The contour lines are **10 µT** apart. The mean field at the center was found to be 3.875 mT. $B_m$ field inhomogeneity over a 22 cm diameter was 1300 ppm.

where *H* is the magnetic flux density and *J* is the current density, from which the static magnetic field B is given as $B = \nabla \times A$ and $H = \mu B$ where *A* is the magnetic vector potential, and *µ* is the permeability of the material. The generated profile of the magnetic field is shown in Figure 1.

**The $B_p$ magnet coil**
In PMRI systems, magnetic field inhomogeneities of up to 30% for the $B_p$ coil are considered acceptable as such non-uniformity leads to changes in the brightness of the image [11]. The brightness can be corrected numerically. The $B_p$ magnet coil was therefore constructed as a simple solenoid coil, which produces relatively large magnetic fields for a given copper mass and provides sufficient field uniformity. The maximum power requirement was set at 500 W. A cylindrical aluminum tube with inner diameter of 30 cm and outer diameter of 30.8 cm was used as the coil supporting structure. Wire size was determined to be 2.6 mm in diameter. The wire length was 1525 m based on power requirements, and coil resistance was calculated to be 5 Ohms. The required number of turns as calculated from wire length and tube diameter would be 1576 turns.

To construct a solenoid coil with field inhomogeneity within 30% over the field of view of 25 cm in length, dimensions were optimized with the COMSOL model by varying the length of the solenoid from 25 cm to 45 cm in steps of 5 cm while the magnetic field strength along the axis was evaluated. It was observed that while magnetic field homogeneity of the solenoid improved with increasing length, magnetic field strength was reduced. The optimized $B_p$ coil was wound as a solenoid of length 40 cm, inner diameter 30 cm, and outer diameter 38 cm. The number of turns was 1188, actual solenoid resistance was measured to be 3.5 Ohms and inductance of 40 mH giving a time constant of 50 ms. The average field strength over the length of the solenoid was 3.7 mT/A.

**Gradient coils**

Gradient coils were based on standard designs. The Z-gradient was made from a Maxwell coil pair of radius 25.53 cm. Coil separation for the z-gradient was 44.72 cm. Maxwell coils with such specifications give a uniform and linear magnetic field to a spherical region defined by a radius of ½$R$ [28]. This is a large enough region for our purpose given the radius of our coils. The transverse gradients were constructed as Golay coils [29] with diameter of 54.5 cm, inner arc separation of 21.25 cm and outer arc separation of 68 cm. Litz wire was used for the gradient coils rather than single stranded copper wire to make it smoother to wind round sharp corners as required in making Golay coils.

**The radio frequency coils**

We used two separate radio frequency (RF) coils – one as the transmit coil and the other as the receiver coil. We considered two major parameters when designing both RF coils – the resonant frequency and the quality factor. Both RF coils were saddle coils with length to diameter ratio of 1.2.

The receiver coil had a diameter of 22 cm and arc angle of approximately 120°. The saddle wire packs comprised 10 turns of a 105-strand 1.29 mm Litz wire. High Q-factor ceramic capacitors with values ranging between 1 to 5 nF were used to tune the coil to frequencies between 100-300 kHz.

The transmit coil had a diameter of 27 cm and an arc angle of approximately 120°. The saddle wire packs had 4 turns of 105 strands Litz wire with nominal diameter of 1.29 mm. Interaction between coils was minimized by orienting the coils in such a way that their magnetic fields were orthogonal to each other.

A SR560 low-noise voltage preamplifier from Stanford Research Systems was used for the receive coil and a home-built RF amplifier based on a OPA 134 Op-amp configuration was used for the transmit coil.

**The radio frequency (RF) shield**

The imaging volume of the PMRI system was partially shielded using an open-ended cylindrical aluminum shield. An aluminum sheet 0.8128 mm thick was formed into a cylinder of 28.5 cm in diameter and 61 cm in length. The skin depth δ of a material is given by [30]

$$|\delta| = \sqrt{\frac{2\rho}{\mu\omega}} \qquad \text{Equation 3}$$

Where ρ is the resistivity of the material, μ is the permeability, and ω is the frequency.

Our system is designed to work at between 100-300 kHz and the skin depth for aluminum at these frequencies is much smaller than the thickness of the shield. Theoretically, electromagnetic interference (EMI) is shielded by absorption and reflection, and the shielding effectiveness (S.E.) of a material is given by [31],

$$S.E\ (dB) = 20\log\frac{\eta_o}{4\eta} + 20\frac{x}{\delta} \qquad \text{Equation 4}$$

where $\eta_o$ is the wave impedance of air and $\eta$ is the wave impedance of the material, $x$ is the material thickness and $\delta$ is the skin depth of the material. The shielding effectiveness for the open ended aluminum shield at our frequency of operation, calculated based on [32] was found to be 34dB. The grounding terminal welded onto the shield was connected to the electrical ground of the building through the power outlet. The eddy currents within the shield must decay faster than coil transients, because with the chosen delay in the π/2 pulse, we do not observe significant effects of eddy currents. A small hole at one end of the shield permitted entry of magnet, gradient and RF wires to the respective coils.

**The magnet coil assembly**
Following the procedures described above, the PRMI system was finally wound by hand and assembled starting with the receiver coil as the inner-most coil, the transmit coil, the $B_p$ coil, the $B_m$, the Z-gradient coil, and X and Y gradients as the outer most coils the as shown in figure 2. The system consists of coils described in the methods section and is mounted on a wooden frame. The overall length of the system is 67 cm and the internal diameter is 22 cm. The system, which includes the coil system and housing, weighs 125kg and costs $18,000 in materials. The electronic components which included power supplies, gradient and signal amplifiers costs roughly $12,000. The control system was implemented in LabView and the cost of developing the LabView code is not included here.

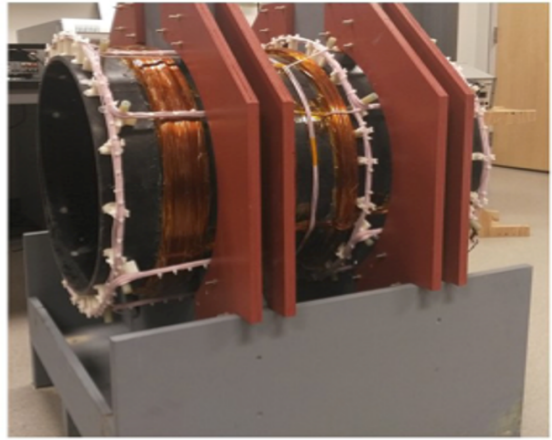

Figure 2: A photograph of the PMRI coil system after construction and assembly, mounted on a wooden frame. Gradient coils are seen as white circular rings and the $B_m$ coils are visible as brown rings

**Results**

After construction of the PMRI system, homogeneity and linearity of the magnetic fields generated by the $B_m$ and gradient coils were evaluated, and imaging experiments were performed.

**The $B_m$ field homogeneity**

The magnetic field inside the PMRI system was mapped using a gaussmeter (DSP 475,

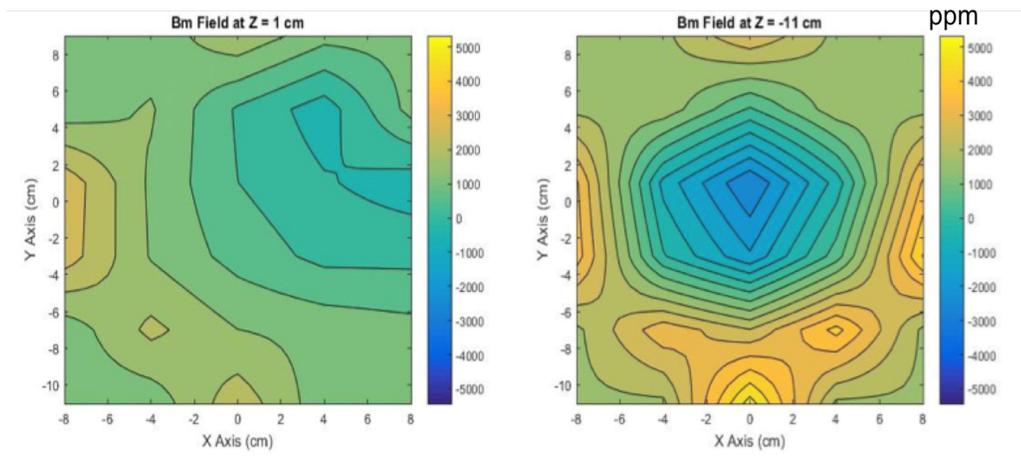

Figure 3: Bm magnetic field homogeneity on the XY plane at z=1cm (left) and z=-11cm (right). The contour lines are 500 ppm. The scale bar is in parts per million (ppm) and the mean field strength is 4 mT.

Lake Shore, USA). The Gaussmeter probe was mounted on the Z-axis of an in-house built 3-axis manipulator capable of adjustments to 2 mm precision. The $B_m$ coil was energized by injecting a current of 4.71 A at 22 V using a DC Power Supply (Xantrex XDL 35-5 T, USA). Measurements were taken at 1 cm intervals over a volume of 16 cm x 16 cm x 16 cm. Results were processed in MATLAB to generate the magnetic field map showing homogeneity of the $B_m$ magnetic field. MATLAB contour plots showing magnetic field homogeneity on the xy-plane at z=1cm and Z= -11cm are shown in figure 3.

**Linearity and uniformity of the gradient magnetic fields**

Similarly, the magnetic field strengths generated by gradient coils were measured using the DSP 475 Gaussmeter guided by the 3-axis manipulator. Each gradient coil was separately energized by supplying a steady current of 2.5 A using the XDL 35-5 T DC Power Supply as above. Results for gradient field uniformity and linearity were analyzed using MATLAB and plotted in figures 4 and 5.

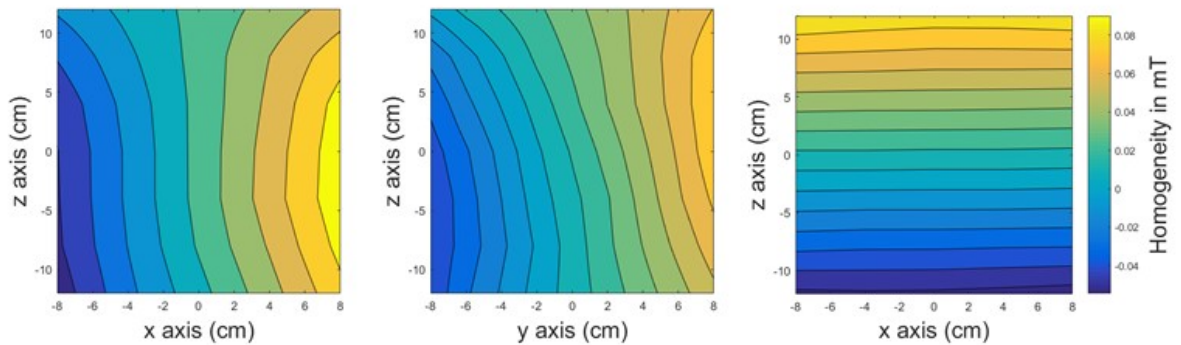

Figure 4: Contour plots of magnetic field uniformity for the X, Y and Z gradients. The contour lines are **10 µT** apart. Curvature of the contour lines indicates the level of field non-uniformity.

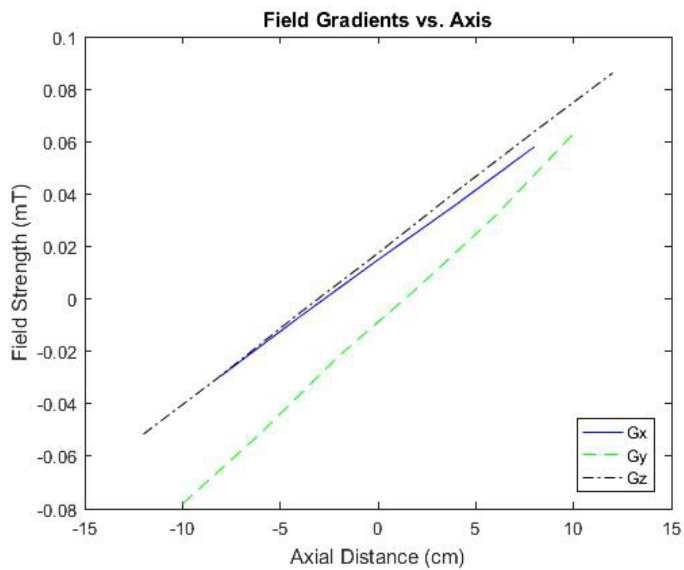

Figure 5: Plots showing linearity of the X-gradient (Gx), Y-gradient (Gy), and Z-gradient (Gz) magnetic fields. The slopes of the lines indicate the strengths of the gradient magnetic field.

**Imaging experiments**

Imaging experiments of phantoms were carried out using the PMRI system. Two separate RF coils – transmit coil and receiver coil were used. They were positioned cylindrically concentric to each other but with orthogonal magnetic fields. Imaging frequency was 113 kHz. The $B_p$ field was 27 mT and the strongest gradient was ~0.7 mT/m. To cover the field of view (FOV) of 22 cm, the RF bandwidth was chosen to be 7kHz.

The pulse sequence used is shown in figure 6 and was controlled by the LabView code. The imaging process started by switching $B_p$ on for 3.5 s followed by an off state of 0.85 s within which image acquisition was done. The long $B_p$ time was to take advantage of the relatively long T1 time of CSF [5]. Similarly, the relatively long time between $B_p$ off and the π/2 pulse was to allow the inductive effects of $B_p$ to fade out.

For the experiment in figure 7, repetition time was 4.35s, number of phase encoding steps was 30 and the number of slices was one. Imaging time was 8.7 minutes. The figure also shows photos of the bell pepper which was imaged. In one of the photos, the bell pepper was cut open to show the internal components which correlate to what is seen in the image.

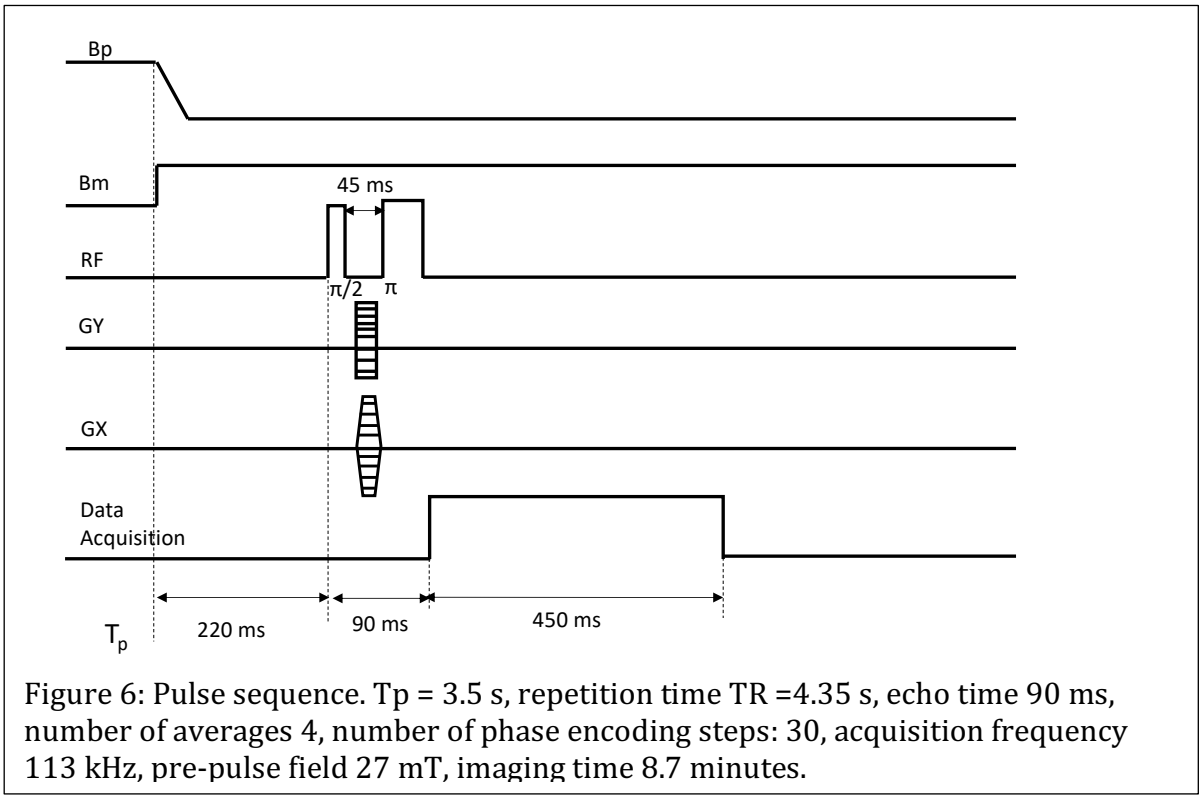

Figure 6: Pulse sequence. Tp = 3.5 s, repetition time TR =4.35 s, echo time 90 ms, number of averages 4, number of phase encoding steps: 30, acquisition frequency 113 kHz, pre-pulse field 27 mT, imaging time 8.7 minutes.

Figure 8 on the other hand shows the image and photo of water bottles, 10mm in diameter and 15mm in length. The bottles were filled with ordinary tap water without any paramagnetic particles. The bottles were imaged using the following parameters: Repetition time of 4.35s, number of phase encoding steps was 26, number of averages was 10, number of slices was one and imaging time was 18.85 minutes.

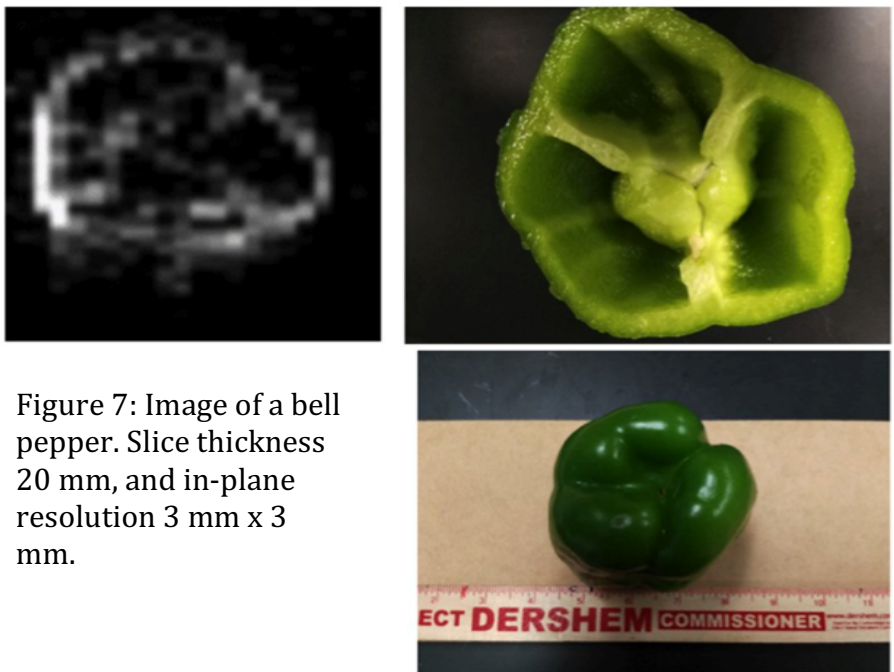

Figure 7: Image of a bell pepper. Slice thickness 20 mm, and in-plane resolution 3 mm x 3 mm.

Figure 9 shows photos and images of a water melon. The water melon was cut into two and the inside flesh scooped out. A water filled balloon was then inserted into the space and the water melon was covered up and wrapped with kapton tape. Imaging of the water melon also followed the pulse sequence diagram shown in figure 7 with repetition time of 4.35 s, number of averages 5, number of phase encoding steps 26 and imaging time of 9.5 minutes.

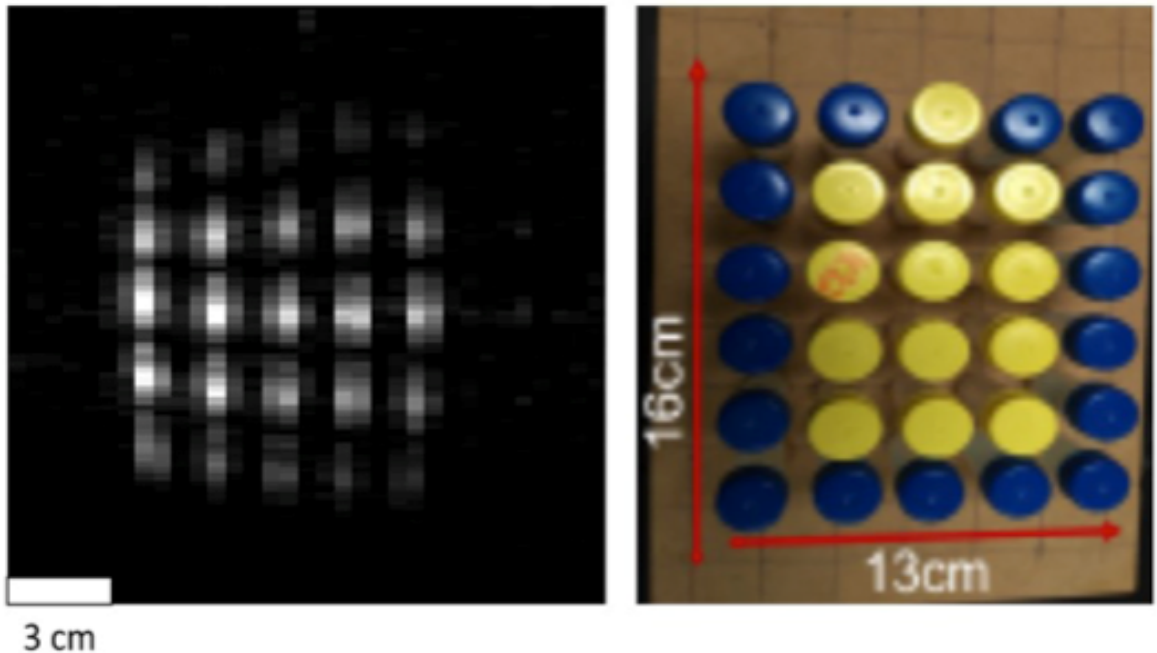

Figure 8: Image of water tubes

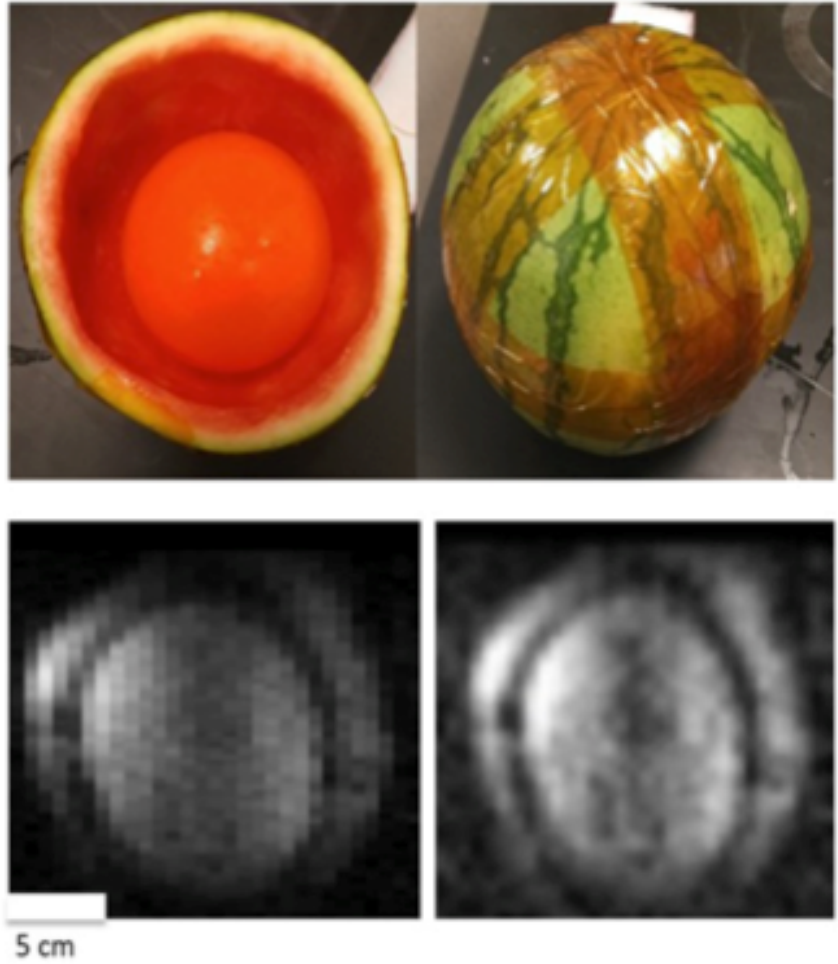

Figure 9: Image of water melon. The bottom image was partially processed using a MATLAB Gaussian filter with standard deviation of 2.

**Discussion**

Using phantom images we have demonstrated that a low-cost PMRI system can reach the imaging resolutions that can be sufficient for some less-demanding clinical applications such as diagnosing hydrocephalus. It is our clinical judgement that hydrocephalus can be managed with an imaging slice thickness of 10mm and in-plane resolution of 3mm x 3mm. We chose a design that can be assembled by hand using widely available materials and off-the-shelf electronics. Operating and maintaining such a system has far lower technical and infrastructure requirements compared to typical commercial imaging systems, which are essential for deploying this technology in many countries in the developing world.

The $B_m$ magnetic field varied up to 5000 ppm within the FOV with average field strength of 4 mT (Figure 3). There is a slight departure from theoretical calculations in terms of the absolute values of magnetic fields measured. Theoretical calculations are that a magnetic field of 0.83 mT/A and therefore a current of 4.721 A would give an average field strength of 3.94 mT. From COMSOL simulations (Figure 1) the mean field at the center was found to be 3.875 mT and field uniformity over the 22 cm DSV was found to be 0.13% (1300ppm). Some deviation could be attributed to the imprecise positioning of coils as a result of hand winding. Kedzia et al. have shown that even slight changes in positioning of coils can cause substantial distortions in the homogeneity of the magnetic field [34]. Additionally, kinks within the coil as a result of hand winding distort the directional flow of current and can create current loops, which further distort the homogeneity of the magnetic field.

However, it is recommended that magnetic inhomogeneities should be smaller than the gradient magnetic field strength per voxel or pixel [22]. Low-field MRI image resolution of 2.5 mm x 3.5 mm has been shown to reveal considerable brain structure [35]. For diagnosis of hydrocephalus, we believe that a more coarse in-plane resolution of 3 mm x 3 mm is adequate, because in the fmajority of cases tissue contrast to reveal brain structure is not pertinent to treatment outcome. The $B_m$ field inhomogeneity in our experiment causes frequency shifts of 12 Hz over a 3 mm pixel. This implies that our gradient strengths should be greater than 12 Hz per pixel, which is easily achieved with our gradient system.

Figure 4 shows contour plots of the *X, Y* and *Z* magnetic field gradients respectively. The contour lines in all cases are 10 µT apart. Magnetic field uniformity is shown by how parallel the contour lines are while linearity is shown by how straight the contour lines are. We observe that the gradient fields are fairly uniform but are linear within a limited range. Linearity of the gradient fields are explored further by the line graphs in Figure 5. Gradient strengths in the *X, Y* and *Z*-directions were calculated from the slopes of the graphs in Figure 5 and found to be 542 mT/m, 697 mT/m and 577 mT/m respectively.
Considering the pixel size of 3 mm, these gradient fields give frequency shifts of 69 Hz/pixel, 89 Hz/pixel and 74 Hz/pixel in the *X, Y* and *Z*-directions respectively. These frequency shifts caused by $B_m$ inhomogeneity is only 17% of the weakest gradient strength and therefore permits our successful imaging in this design.

From the images acquired (Figures 7, 8, and 9), distortions that are a reflection of the $B_m$ field inhomogeneities and non-linearities can be inferred. In the larger sample images (Figures 7 and 9), a loss of signal can be seen towards the edges of the images. This could be due to a number of factors. The RF transmitter coil is comparable in length to the dimensions of the samples being imaged. Ginsberg and Melchner found that a saddle coil gives the most uniform field when the length of the coil is twice its diameter [36]. When dealing with large volume coils, it becomes difficult to adhere to these dimensions as the saddles could be unnecessarily long. Our saddle coil has a length to diameter ratio of 1.2. The RF magnetic field is therefore not uniform throughout the imaging volume. Simulations suggest that the usable volume of a saddle RF coil is that region where the magnetic field is uniform to within 15% [37]. RF field inhomogeneities towards the edges of the coil lead to loss of signal. Similar to the transmitter coil, the receive coil has dimensions comparable to those of the samples. Additionally, the gradient magnetic fields are more uniform at the center of the coil system with non-uniformities at the outer edges as seen from figures 7, 8 and 9. This can result in spin dephasing at the edges. We also see a brightening of the images on the left side, which could be caused by proximity of the samples to the RF shield inserted into the imaging volume. The RF shield can produce non-uniform concentration of the RF signal causing brightness that would be visible in the images. The distortions in the image of the water bottles could be caused by gradient non-linearities as the contour plots in Figure 4.

Furthermore, magnetic field instabilities caused by the $B_p$ switching can cause spin dephasing and these can be more pronounced at the edges of the samples where the non-uniformity may be greater. This dephasing causes signal loss that can affect image contrast making it difficult to define boundaries between different structures.

The images acquired were of samples with high water content. Water is known to have long longitudinal relaxation times (T1) [38] and therefore our delay time of 220 ms before the application of the $\pi/2$ pulse was still appropriate. However, human tissues have relatively short T1 relaxation times close to 200 ms especially at low magnetic fields [14]. We are working towards reducing the delay time by improving $B_p$ switching. Nevertheless, imaging can be done for maximum contrast of CSF, which is particularly relevant to hydrocephalus diagnostics.

**Conclusions**

Coil designs and construction used in the PMRI system described here were based on general procedures proposed and described by many researchers [22],[23], [28],[30],[39]. The fact that we were able to acquire usable images with $B_m$ magnetic field inhomogeneities of up to 5000 ppm in an unshielded room is testament to the robustness of the PMRI technology and an indication of its suitability for applications in developing countries. The signal to noise ratio (SNR) of the system can be greatly improved by increasing the $B_m$ magnetic field until coil noise is dominated by sample (tissue) Johnson noise [11]. This however needs to be done careful as increasing the strength of $B_m$ would substantially increase the power requirement of the system. Despite the imperfections seen in such a simple device, the images acquired show the potential functionality of the system for our intended first application in infant hydrocephalus. The fact that the system was assembled by hand shows the robustness of its ease of construction, which make it straightforward to replicate and scale in developing countries. The system was operated at a $B_m$ of 2.66 mT and a $B_p$ of 27 mT with power consumption of 140W and 410W respectively. At this low power, the in-plane resolution of 3 mm x 3 mm is adequate for treatment decisions in hydrocephalic cases where the primary need is to distinguish between CSF and brain tissue. For deployment in Africa, we have embarked on a program to automatically compensate for field imperfections, optimize transmit-receive coil specifications, achieve adequate thermal and electrical safety specifications, and optimize a variety of image acquisition algorithms to further improve spatial resolution and contrast. Redesigning such a system for adult head imaging will be a straightforward extension of this effort.

## Acknowledgements

This work was funded by the Endowment funds of Harvey F. Brush at Penn State University, and US NIH Grant 5DP1HD086071.

## AUTHORS CONTRIBUTIONS

**Johnes Obungoloch:**
Responsible for winding magnet coils and carrying out imaging experiments

**Joshua Harper:**
Responsible for solid work Computer Aided Designs (CAD) of the magnet and coil systems, and wasresponsible for image analysis

**Steve Consevage:**
Responsible for theoretical analysis of magnetic field uniformity for the designs

**Srinivas Tadigadapa:**
Participated in designing building radio frequency amplifiers used in this project.

**Thomas Neuberger:**
Participated in radio frequency tuning of coils, and contributed to image analysis.

**Igor Savukov:**
Consulted on all aspects of device design, wrote the labview program used in imaging, and helped carry out imaging experiments.

**Steven J. Schiff:**
Overall supervisor for the concept, design, implementation, and analysis for project.

March 28, 2018 arXiv:1803.09075 [physics.med-ph]**References**

1. K. T. Kahle, Kulkarni, A. V., Limbrick, D. D., Jr., and Warf, B. C. (2016). Hydrocephalus in children. *Lancet, 387*(10020), 788-799.
2. B. C. Warf, (2005). Hydrocephalus in Uganda: the predominance of infectious origin and primary management with endoscopic third ventriculostomy. *J Neurosurg, 102*(1 Suppl), 1-15.
3. B. C. Warf *et al.*, "Costs and benefits of neurosurgical intervention for infant hydrocephalus in sub-Saharan Africa.," *J. Neurosurg. Pediatr.*, vol. 8, no. 5, pp. 509–21, 2011.
4. A. V. Kulkarni, Schiff, S. J., Mbabazi-Kabachelor, E., Mugamba, J., Ssenyonga, P., Donnelly, R., Levenbach, J., Monga, V., Peterson, M., MacDonald, M., Cherukuri, V., and Warf, B. C. (2017). Endoscopic Treatment versus Shunting for Infant Hydrocephalus in Uganda. *N Engl J Med, 377*(25), 2456-2464.
5. W. D. Rooney *et al.*, "Magnetic Field and Tissue Dependencies of Human Brain Longitudinal 1 H 2 O Relaxation in Vivo," vol. 318, pp. 308–318, 2007.
6. H.-M. Klein, *Clinical Low Field Strength Magnetic Resonance Imaging: A Practical Guide to Accessible MRI.* New York: Springer, 2016.
7. W. E. Muhogora *et al.*, "Paediatric CT examinations in 19 developing countries: frequency and radiation dose.," *Radiat Prot Dosim.*, vol. 140, no. 1, pp. 49–58, 2010.
8. D. J. Brenner and E. J. Hall, "Computed tomography--an increasing source of radiation exposure.," *N. Engl. J. Med.*, vol. 357, no. 22, pp. 2277–2284, 2007.
9. R. E. Sepponen, J. T. Sipponen, and A. Sivula, "Low Field (0.02 T) Nuclear Magnetic Resonance Imaging of the Brain," *J Comput Assist Tomogr.*, vol. 9, no. 2, pp. 247–241, 1985.
10. G. C. d. Nascimento, Engelsberg, M., and Souza, R. E. d. (1992). Digital NMR imaging system for ultralow magnetic fields. *Measurement Science and Technology, 3*(4), 370-374.
11. Macovski, A., and Conolly, S. (1993). Novel approaches to low-cost MRI. *Magnetic Resonance in Medicine, 30*(2), 221-230.
12. N. I. Matter, G. C. Scott, T. Grafendorfer, A. Macovski, and S. M. Conolly, "Rapid polarizing field cycling in magnetic resonance imaging," *IEEE Trans. Med. Imaging*, vol. 25, no. 1, pp. 84–93, 2006.
13. Savukov, I., Karaulanov, T., Castro, A., Volegov, P., Matlashov, A., Urbatis, A., Gomez, J., and Espy, M. (2011). Non-cryogenic anatomical imaging in ultra-low field regime: Hand MRI demonstration. *JOURNAL OF MAGNETIC RESONANCE, 211*, 1-23. doi: 10.1016/j.jmr.2011.05.011
14. M. I. Savukov and T. Karaulanov, "Magnetic-resonance imaging of the human brain with an atomic magnetometer.," *Appl. Phys. Lett.*, vol. 103, pp. 1–4, 2013.
15. S. Lother, S. J. Schiff, T. Neuberger, P. M. Jakob, and F. Fidler, "Design of a mobile, homogeneous, and efficient electromagnet with a large field of view for neonatal low-field MRI," *Magn. Reson. Mater. Physics, Biol. Med.*, pp. 1–8,
Page 19 of 21

March 28, 2018    arXiv:1803.09075 [physics.med-ph]

2016.

16. M. Sarracanie, C. D. LaPierre, N. Salameh, D. E. J. Waddington, T. Witzel, and M. S. Rosen, "Low-cost high-performance MRI," *Sci. Rep.*, vol. 5, 2015.
17. K. Halbach, "Design of permanent multipole magnets with oriented rare earth cobalt material," *Nucl. Instruments Methods*, vol. 169, pp. 1–10, 1980.
18. T. Kimura *et al.*, "Development of a mobile magnetic resonance imaging system for outdoor tree measurements," *Rev. Sci. Instrum.*, vol. 82, no. 5, 2011.
19. K. Kose and T. Haishi, "High resolution {NMR} imaging using a high field yokeless permanent magnet," *Magn Reson Med Sci.*, vol. 10, pp. 159–167, 2011.
20. C. Z. Cooley *et al.*, "Two-dimensional imaging in a lightweight portable MRI scanner without gradient coils," *Magn. Reson. Med.*, vol. 73, no. 2, pp. 872–883, 2015.
21. B. Peter and F. Casanova, "Hardware developments: Halbach magnet arrays," in *Mobile NMR and MRI : Developments and Applications*, 2015, pp. 133–154.
22. P. Morgan, S. Conolly, G. Scott, and A. Macovski, "A readout magnet for prepolarized MRI," *Magn. Reson. Med.*, vol. 36, no. 4, pp. 527–536, 1996.
23. J. L. Kirschvink, "Uniform Magnetic-Fields and Double-Wrapped Coil Systems - Improved Techniques for the Design of Bioelectromagnetic Experiments," *Bioelectromagnetics*, vol. 13, no. 5, pp. 401–411, 1992.
24. R. Merritt, C. Purcell, and G. Stroink, "Uniform magnetic field produced by three, four, and five square coils," *Rev. Sci. Instrum.*, vol. 54, no. 7, pp. 879–882, 1983.
25. M. Rubens, Sidney, "Cube-Surface Coil for Producing a Uniform Magnetic Field," *Rev. Sci. Instrum.*, vol. 16, no. 9, pp. 243–245, 1945.
26. G. Gottardi, P. Mesirca, C. Agostini, D. Remondin, and F. Bersani, "A Four Coil Exposure System (Tetracoil) Producing a Highly Uniform Magnetic Field," *Bioelectromagnetics*, vol. 24, no. 2, pp. 125–133, 2003.
27. Wire industries MWS, "World ' s Largest Selection of Specialty Magnet Wire." [Online]. Available: http://www.mwswire.com/pdf_files/mws_tech_book/techbook2016.pdf.
28. T. S. . Hidalgo, "Theory of Gradient CoilDesign Methods forMagnetic ResonanceImaging," *Concepts Magn. Reson. Part A*, vol. Vol. 36A, no. 4, pp. 223–242, 2010.
29. M. J. E. Golay, "Field homogenizing coils for nuclear spin resonance instrumentation," *Rev. Sci. Instrum.*, vol. 29, no. 4, pp. 313–315, 1958.
30. D. I. Hoult, "The NMR receiver: A description and analysis of design," *Prog. Nucl. Magn. Reson. Spectrosc.*, vol. 12, no. 1, pp. 41–77, 1978.
31. O. W. Henry, *Electromagnetic Compatibility Engineering*. New York: John Wiley & Sons, 2009.
32. R. B. Schulz and W. L. W. Is, "RF Shielding Design," vol. 4, no. 6, pp. 168–175, 1968.
33. Department Of Defense And W. D. 20301, "Military Handbook: Grounding, Bonding, And Shielding For Electronic Equipments And Facilities," Washington,DC, 1987.
34. P. Kedzia, T. Czechowski, M. Baranowski, J. Jurga, and E. Szcześniak, "Analysis of Uniformity of Magnetic Field Generated by the Two-Pair Coil System," *Appl. Magn. Reson.*, vol. 44, no. 5, pp. 605–618, 2013.

35. M. Sarracanie *et al.*, "Low-Cost High-Performance MRI," *Sci. Rep.*, vol. 5, p. 15177, 2015.
36. D. M. Ginsberg and M. J. Melchner, "Optimum geometry of saddle shaped coils for generating a uniform magnetic field," *Rev. Sci. Instrum.*, vol. 41, no. 1, pp. 122–123, 1970.
37. A. Samila, "Simulation of Magnetic Field Topology in a Saddle-Shaped Coil of Nuclear Quadrupole Resonance Spectrometer," *Prog. Electromagn. Res. Lett.*, vol. 56, pp. 67–73, 2015.
38. W. S. Hinshaw, P. A. Bottomley, and G. N. Holland, "Radiographic thin-section image of the human wrist by nuclear magnetic resonance," *Nature*, vol. 270, no. 5639, pp. 722–723, 1977.
39. S. A. Rashid, B. S. Amiruddin, and T. H. Chew, "Magnetic field simulation of Golay coil," *J. Fundam. Sci.*, pp. 353–361, 2008.